\documentclass[12pt]{iopart}

\usepackage{graphicx}
\usepackage[usenames]{color}
 
\begin{document}


\title{Stable matter-wave  soliton  in the vortex core of a uniform condensate}

\author{ S. K. Adhikari\footnote{adhikari@ift.unesp.br; URL: http://www.ift.unesp.br/users/adhikari}
} 
\address{
Instituto de F\'{\i}sica Te\'orica, UNESP - Universidade Estadual Paulista, 01.140-070 S\~ao Paulo, S\~ao Paulo, Brazil
} 

\begin{abstract}

We demonstrate a   stable, mobile, dipolar or nondipolar three-dimensional 
matter-wave soliton
in  the vortex core of a uniform nondipolar  condensate.  
All intra- and inter-species contact interactions can be repulsive for   a strongly  dipolar soliton.
For a weakly dipolar or nondipolar soliton, the intra-species contact interaction in the soliton 
should be attractive for the formation of a compact soliton. 
The   soliton can  propagate with a constant velocity along the vortex core   
without any  deformation.  
Two such solitons   undergo a quasi-elastic collision at medium velocities.  
We illustrate the findings using realistic interactions in a mean-field model of  
 binary $^{87}$Rb-$^{85}$Rb and $^{87}$Rb-$^{164}$Dy systems.

\end{abstract}


\maketitle

  \section{Introduction}
 
A bright soliton is a self-bound    object that 
travels at a constant velocity  without deformation
in one dimension (1D), due to a cancellation of 
nonlinear attraction and defocusing forces. 
{An 1D dark soliton is a dip in  uniform density, which also moves with a constant velocity maintaining its shape.} 1D Solitons   have been observed   in nonlinear optics \cite{book}, and in Bose-Einstein condensate (BEC) \cite{rmp}.  
Experimentally, bright matter-wave solitons   were created in  
 BECs of
$^7$Li \cite{1}
 and
$^{85}$Rb atoms \cite{3}. Dark solitons were also observed in BECs  of $^{87}$Rb \cite{dark}  and 
$^{23}$Na \cite{dark1}.
The 1D set up is obtained by putting confining traps in   directions perpendicular to the motion of the soliton.   
However,   a three-dimensional (3D)  trap-less soliton
cannot be formed 
for a cubic  nonlinearity, generally encountered in BEC and nonlinear optics,
 due to collapse \cite{sptem1}.   
The collapse can be stopped in a weaker
saturable \cite{4-5}, cubic-quintic \cite{3dvor} or quadratic \cite{6-8} nonlinearity, 
or by  an application of  nonlinearity and/or dispersion  management \cite{9}.
In nonlinear optics, 1D temporal solitons \cite{1dopttem} as well as  lattice solitons in arrays of nonlinear optical wave guide     in 1D \cite{1doptsp} and in two dimensions (2D) \cite{2dopt} and 3D 
\cite{3dopt}, with  modified dynamics/nonlinearity,
have been observed. 

Here we 
demonstrate the formation of a  {\it trap-less}   matter-wave soliton in the core of a quantized vortex of a uniform nondipolar BEC, which we call a binary nondipolar  vortex-soliton, which is a 3D analogue of an 1D dark-bright soliton \cite{book} studied previously. 
These solitons  are shown to be stable and execute steady oscillation for very long time under a small perturbation.  
 In the case of a 3D vortex-soliton
all interactions can be repulsive except the intra-species interaction in the soliton. 
The soliton can swim freely with a constant velocity  along the vortex core.
  Because of the strong localization 
of the soliton due to inter-species contact repulsion, the soliton can move without visible deformation,
The collision between two integrable 1D solitons is truly elastic \cite{book,rmp}.
However,   at medium velocities the collision between two solitons  
is found to be quasi-elastic without visible deformation.  
 As no modification of the nonlinear interactions is suggested such a trap-less soliton can be realized in a laboratory. 
 

The observation   of dipolar  BECs of $^{164}$Dy \cite{ExpDy}, $^{168}$Er \cite{ExpEr} and 
  $^{52}$Cr \cite{cr}  
has initiated  studies   of  new types of BEC solitons. For example, one can have a 
 dipolar BEC soliton for a repulsive 
contact interaction \cite{1D},    in 2D \cite{2D}
or  on optical-lattice potentials \cite{ol2D}.
These new dipolar solitons were  possible due to the peculiar nature of dipolar interaction.
The dipolar BEC solitons of a large number of atoms
stabilized  by a long-range dipolar  attraction, 
could be  robust and less vulnerable to collapse   in the presence of a  short-range 
contact repulsion \cite{1D,stablesol}. Hence, we also consider a dipolar soliton in a nondipolar vortex core, which we call a dipolar vortex-soliton.  
 Actually, a strong short-range inter-species contact repulsion between  the atoms of the matter-wave soliton and the atoms of the vortex    localizes the soliton in 3D.

In the present investigation we use a mean-field model described in section \ref{II}. 
First we present an 1D model in section  \ref{IIA} useful for an analytic understanding of the 
formation of a dark-bright soliton. The 3D mean-field model is presented in section  \ref{IIB}. In section \ref{III} we present
numerical results for the formation of a binary vortex-soliton. A summary of our findings is given in section \ref{IV}.

\section{Mean-field model}

\label{II}

\subsection{One Dimension}

\label{IIA}
 
\begin{figure}[!t]

\begin{center}
\includegraphics[width=\linewidth]{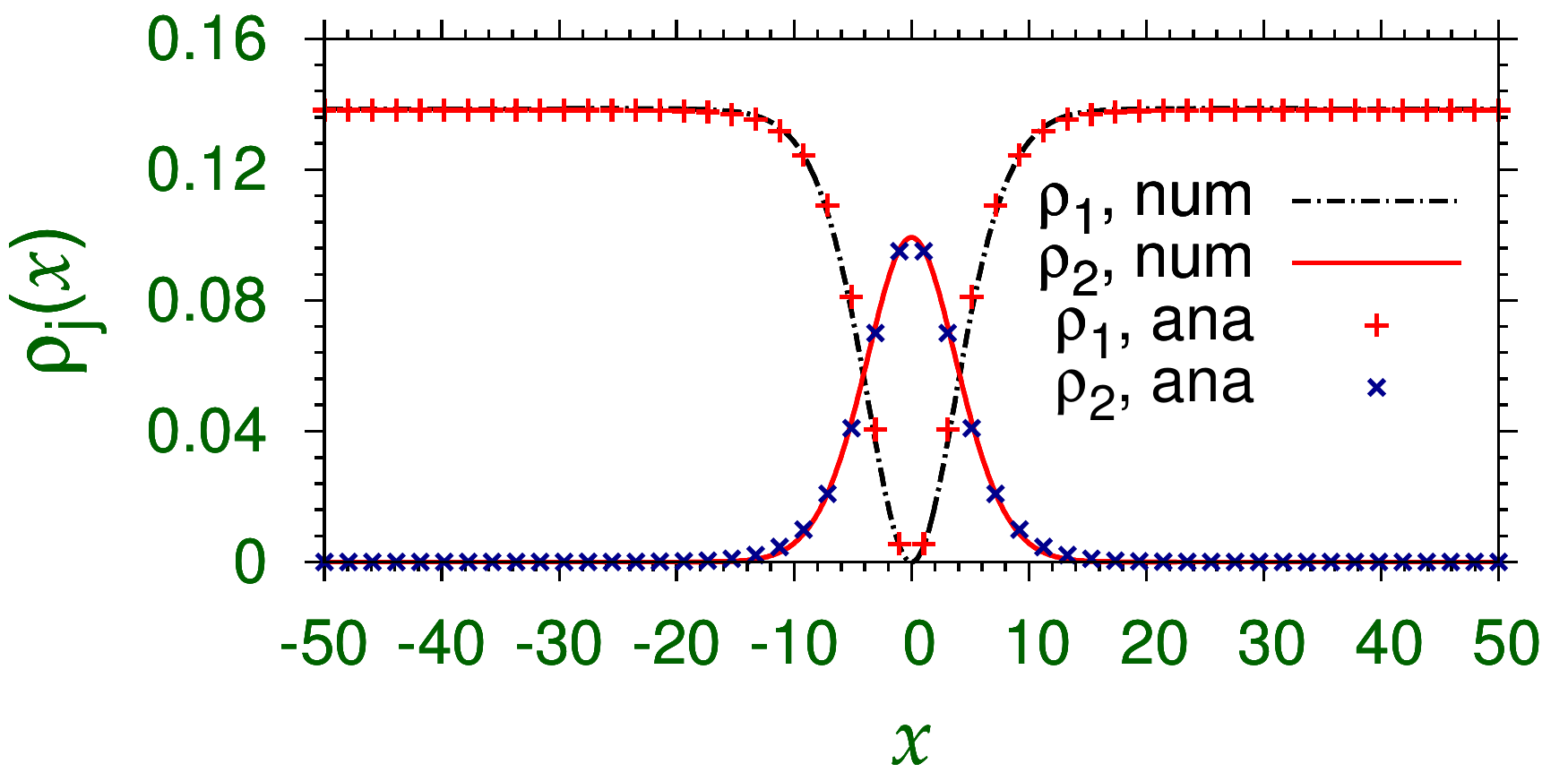}

\caption{ (Color online)  
Densities $\rho_j(x)=|\psi_j(x)|^2$ from analytic (ana) model (\ref{xx1}) and (\ref{xx2})
with $\alpha=0.198, \beta=\sqrt{a^2+a/2}=.372$ 
compared 
with those of numerical (num) simulation by imaginary-time propagation in 
  the space domain $x=\pm L=\pm 50.$ The normalization of the wave functions are $\int_{x=-L}^L \rho_1(x)dx =12.42$ and 
$\int_{x=-L}^L \rho_2(x)dx =1.$ 
}\label{fig0} \end{center}

\end{figure}

    A vortex in a   uniform BEC bears similarity with 
an 1D dark soliton in having a hole along the axial $z$ direction and is often called a 3D dark soliton \cite{book}. 
A binary vortex-soliton is the 3D analogue of the well-known  1D dark-bright soliton.
Hence, 
to understand how  a  3D vortex-soliton can appear,  we consider the following integrable binary 1D
dark-bright soliton  model  of atoms of same mass
$m$,  
same inter- and intra-species scattering lengths $a$ and same number of atoms $N$ \cite{luca}
\begin{equation}
 i\hbar \frac{\partial \phi_j(x,t)}{\partial t}=
{\Big [}  -\frac{\hbar^2}{2m }\frac{\partial^2}{\partial x^2}
+ g\sum _{i=1}^2\vert \phi_i \vert^2
{\Big ]}  \phi_j(x,t),
\end{equation}
where $i,j=1,2,$ $g=2a\hbar^2 N/(md_\rho^2)$, $d_\rho=\sqrt{\hbar/(m\omega)}$, where $\omega$
is the frequency of 
 a strong harmonic trap in the binary mixture in the transverse directions.  
Scaling the wave functions by  $|\psi_i|^2=g|\phi_i|^2$ we obtain 
\begin{equation}\label{model}
 i \frac{\partial \psi_j(x,t)}{\partial t}={\Big [}  -\frac{1}{2 }\frac{\partial^2}{\partial x^2}
+\sum_i\vert \psi_i \vert^2 {\Big ]}  \psi_j(x,t), 
\end{equation}
in   units $\hbar=m=1.$  Equation (\ref{model}) with {\it all-repulsive} interactions 
has the analytic dark-bright soliton \cite{book} 
\begin{eqnarray} \label{x1}
\psi_1(x,t)= && \beta \tanh[\alpha(x-vt)]e^{ivx-i(v^2/2+\beta^2) t},\\
\psi_2(x,t)=&&\gamma\mathrm{sech}[\alpha(x-vt)]  e^{ivx+i[(\alpha^2-v^2)/2-\beta^2]t},
\label{x2}
\end{eqnarray}
where $\alpha$ and $\beta$ $(\beta>\alpha)$ are  constants which control the intensity and width of the solitons,  $\gamma=\sqrt{\beta^2-\alpha^2} $ and 
$v$ is the  velocity. The appearance of  bright soliton (\ref{x2}) in the 
all-repulsive equation (\ref{model}) 
is counterintuitive and is possible due to the coupled dark soliton (\ref{x1}). 
Without losing generality 
we impose the 
normalization    
$\int |\psi_2(x,t)|^2dx =1$, yielding $  \beta=\sqrt{\alpha^2+\alpha/2}$. 
For $v=0$ the solutions have the  form
\begin{eqnarray} \label{xx1}
\psi_1(x,t)= && \sqrt{\alpha^2+\alpha/2} \tanh(\alpha x)\exp[-i(\alpha^2+\alpha/2) t],\\
\psi_2(x,t)=&&\sqrt{\alpha/2}  \mathrm{sech}(\alpha x)  
 \exp[-i(\alpha^2+\alpha)t/2]. \label{xx2}
\end{eqnarray}  
To solve  (\ref{model}) numerically  
the simulation was performed in a 1D box in the domain $x=\pm 50$. 
In figure \ref{fig0} we plot the numerical   matter-wave densities of the dark-bright soliton    and  the analytic results (\ref{xx1})
and (\ref{xx2}).
The bright soliton sits in the central hole of the dark soliton and the inter-species repulsion
between the (outer) dark   and (inner) bright solitons  confines the latter.     Similarly, the soliton of a vortex-soliton
can be confined in the radial $x-y$ plane by the inter-species repulsion between the vortex and   soliton. The confinement along the $z$ direction is obtained 
by the intra-species dipolar attraction with the dipoles polarized along   $z$ axis and/or by intra-species contact attraction. 
   
\subsection{Three Dimensions}

\label{IIB}

We will consider a matter-wave vortex-soliton  in the form of a   vector
soliton  without any trapping potential and present the binary BEC model 
appropriate for this study. 
The first component  ($j=1$), with the vortex, is   nondipolar and the 
second component ($j=2$), with the   soliton, 
  can be dipolar  or nondipolar. As in the 1D dark-bright soliton, the  intra-species repulsion  
prevents  the matter-wave soliton from escaping radially. The dipolar interaction or intra-species 
contact attraction prevents the soliton from escaping in the axial  direction.
The 
mass, number of atoms, and scattering length for the two species  
are $m_j, N_j, 
 a_j,$ respectively. 
  The intra- ($V_{j}$)
and inter-species ($V_{12}$)
interactions 
for   atoms at  $\bf r$ and $\bf r'$ are \cite{mfb2}
\begin{eqnarray}\label{intrapot} 
V_1({\bf R})= \frac{4\pi\hbar^2 a_1 \delta({\bf R})}{m_1}, \quad 
V_{12}({\bf R})=\frac{ 2\pi\hbar^2 a_{12} \delta({\bf R})}{m_R},\\
V_2({\bf R})= \frac{3
a_{\mathrm {dd}}\hbar^2V_{\mathrm {dd}}({\mathbf R})}{m_2}+\frac{4\pi \hbar^2 a_2 \delta({\bf R})}{m_2}
,
\label{interpot} 
     \end{eqnarray}
 where $\bf R = (r-r'),$
 $  a_{\mathrm {dd}}=$
${\mu_0  \mu^2m_2}/{(12\pi \hbar ^2 )   }$,
$V_{\mathrm {dd}}({\mathbf R})=$ $({1-3\cos^2 \theta})/{{\bf R}^3}$, the reduced mass 
$m_R=m_1m_2/(m_1+m_2)$,
$a_{12}$ is the inter-species scattering length, $a_{{\mathrm{dd}}}$ is a dipolar length to measure the strength of dipolar interaction,
 $\mu_0$ is the permeability of free space, $\mu$ is the magnetic moment of each atom in the dipolar soliton, 
$\theta$ is the angle made by the vector ${\bf R}$ with the  
polarization $z$ direction.
The dimensionless mean-field Gross-Pitaevskii (GP)  equations for the {\it trap-less} binary mixture 
are   \cite{mfb2}
\begin{eqnarray}& \,
 i \frac{\partial \phi_1({\bf r},t)}{\partial t}=
{\Big [}  -\frac{\nabla^2}{2 }+ g_1 \vert \phi_1 \vert^2
+ g_{12} \vert \phi_2 \vert^2
{\Big ]}  \phi_1({\bf r},t),
\label{eq3}
\\
& \,
{ i} \frac{\partial \phi_2({\bf r},t)}{\partial t}={\Big [}  
-m_{12} \frac{\nabla^2}{2}
+ g_2 \vert \phi_2 \vert^2 
+ g_{21} \vert \phi_1 \vert^2  
\nonumber \\  &  \,
+ g_{\mathrm {dd}}
\int V_{\mathrm {dd}}({\mathbf R})\vert\phi_2({\mathbf r'},t)
\vert^2 d{\mathbf r}' 
{\Big ]}  \phi_2({\bf r},t),
\label{eq4}
\end{eqnarray}
where
$m_{12}={m_1}/{m_2},$
$g_1=4\pi a_1 N_1,$
$g_2= 4\pi a_2 N_2 m_{12},$
$g_{12}={2\pi m_1} a_{12} N_2/m_R,$
$g_{21}={2\pi m_1} a_{12} N_1/m_R,$
$g_{\mathrm {dd}}= 3N_2m_{12} a_{\mathrm {dd}}/m_R.$
In  (\ref{eq3}) and (\ref{eq4}), length is expressed in units of 
a  scale   $l$, probability density 
$|\phi_j|^2$ in units of $l^{-3}$,   energy in units of $\hbar^2/(2m_1l^2)$
and time in units of $ 
t_0=2m_1l^2/\hbar$.

  To find a stationary quantized vortex of angular momentum  ${\cal L}$ along  $z$ axis 
in component 1,  also called a dark
soliton with circular symmetry,
 we look for axially-symmetric solution $\Phi_1 ({\bf r},t)$
in $x-y$ plane:  
$\phi_1({\bf r},t) \equiv \Phi_1 ({\bf r},t)e^ {i{\cal L}\varphi} $,
where $\varphi$ is the azimuthal angle  and  $\Phi_1({\bf r},t)$ satisfies \cite{book}
\begin{eqnarray}& \,
 i \frac{\partial \Phi_1({\bf r},t)}{\partial t}= 
{\Big [}  -\frac{1}{2}\Big( \frac{\partial^2}{\partial z^2 }+\frac{\partial^2}{\partial x^2 }+\frac{\partial^2}{\partial y^2 }\Big)\nonumber \\ &+{{\cal L}^2}/{(x^2+y^2)}+ g_1 \vert \Phi_1 \vert^2
+ g_{12} \vert \phi_2 \vert^2
{\Big ]}  \Phi_1({\bf r},t),
\label{eq5}
\end{eqnarray}
 with the boundary conditions
$\Phi_1 (x=0,y=0,z) = 0,
\Phi_1 (x\to \infty,y\to \infty, z) =$ constant. The same on the 1D dark soliton 
(\ref{x1}) are very similar: $\psi_1(x=0)=0, \psi_1(x\to \infty)=$ constant.
For a matter-wave soliton of component 2 in the vortex core of component 1 we solve the axially-symmetric equations (\ref{eq5}) and (\ref{eq4}).  We consider a vortex of unit circulation  ${\cal L}= 1$.

\begin{figure}[!t]

\begin{center}
\includegraphics[width=.49\linewidth,clip]{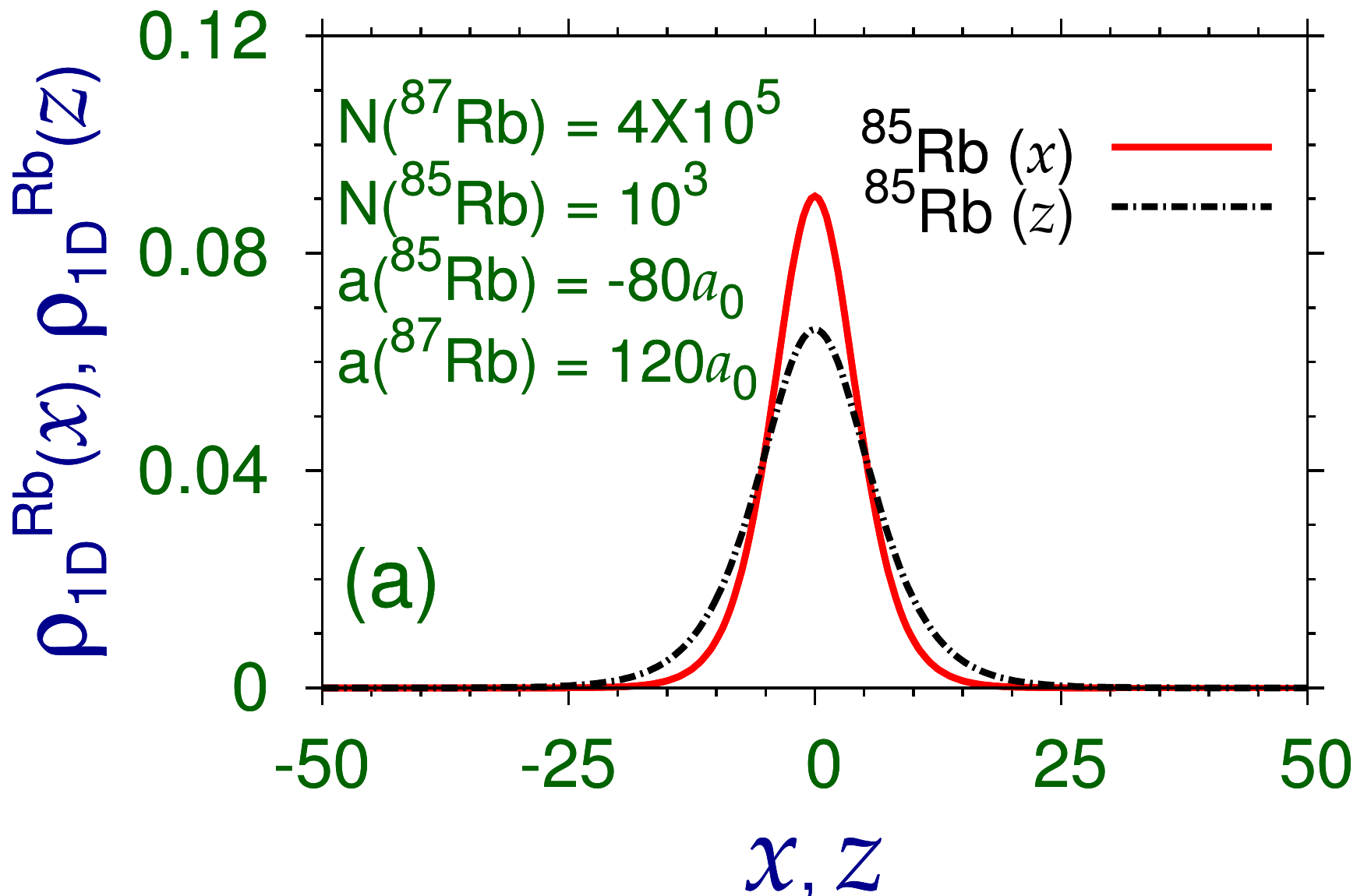}
\includegraphics[width=.49\linewidth,clip]{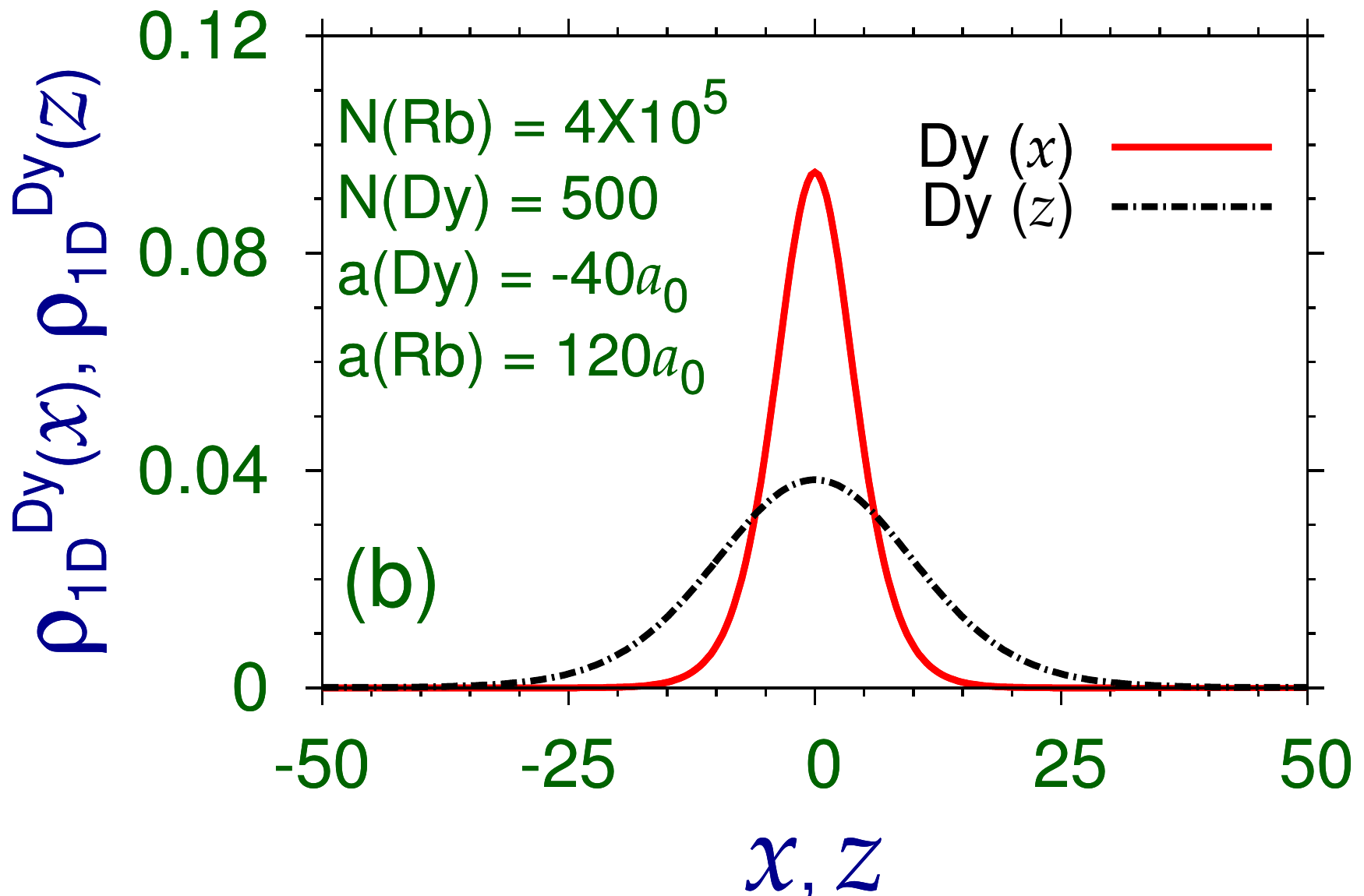}
\includegraphics[width=.49\linewidth,clip]{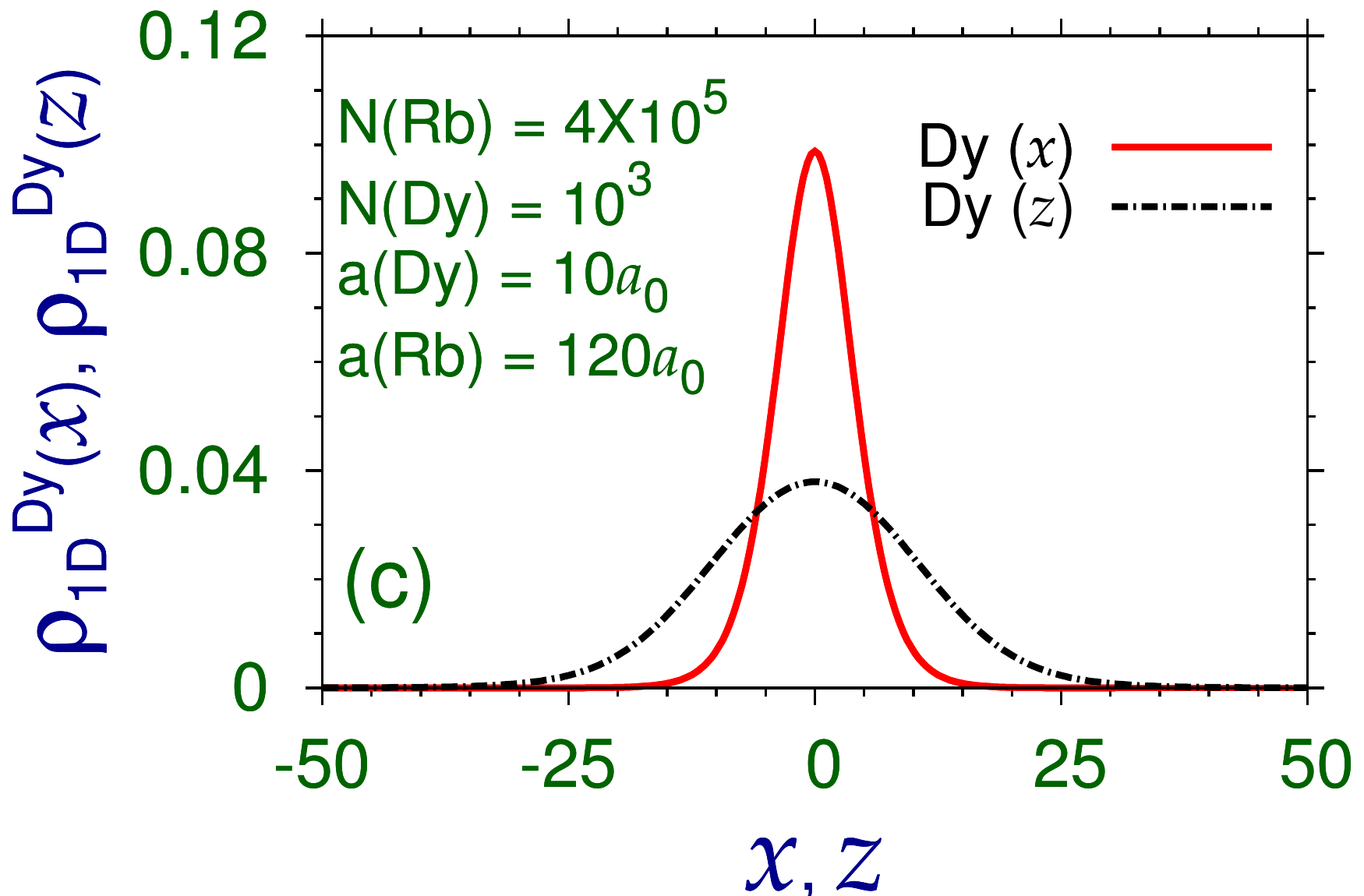}
\includegraphics[width=.49\linewidth,clip]{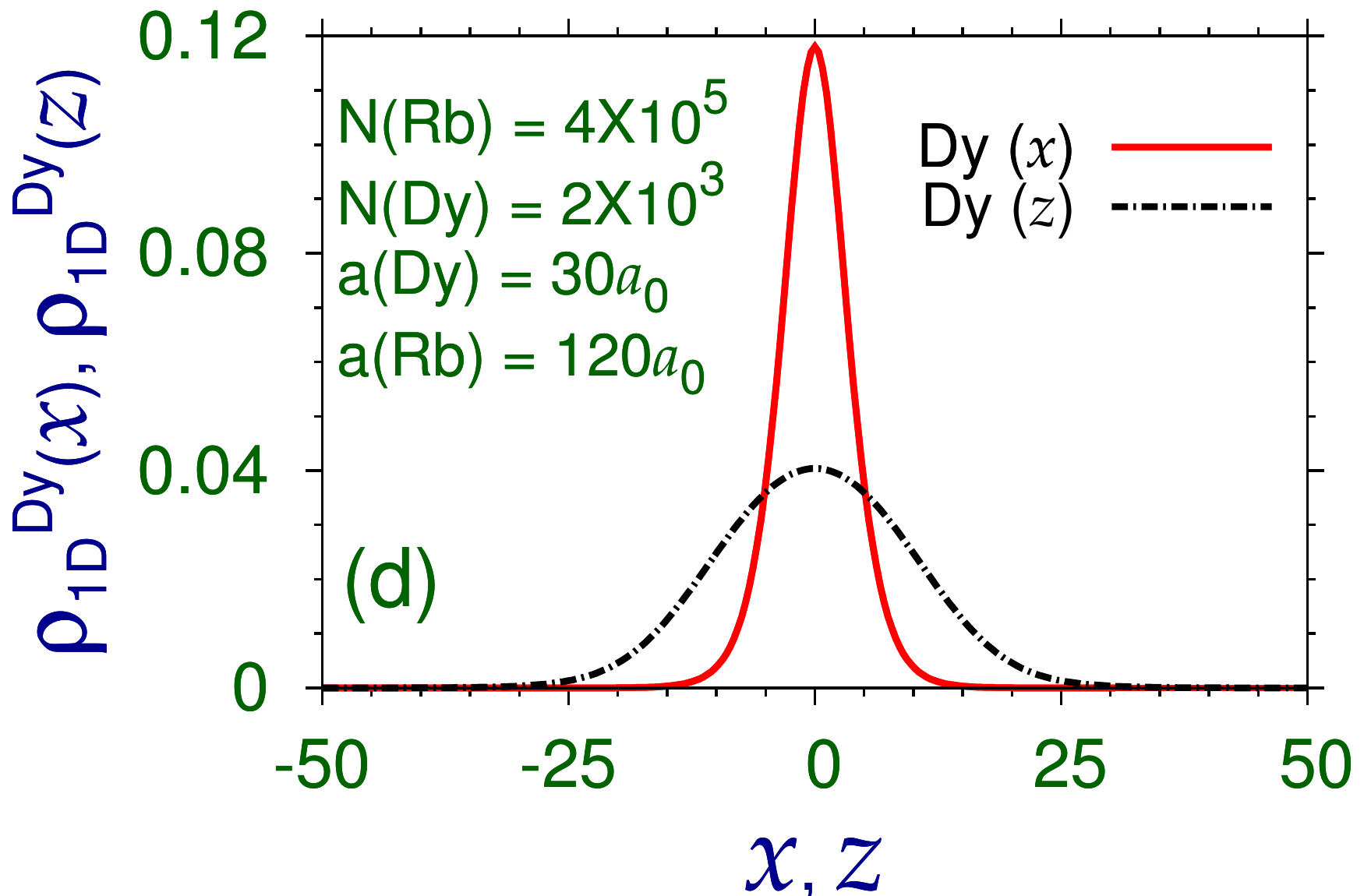}
\includegraphics[width=.49\linewidth,clip]{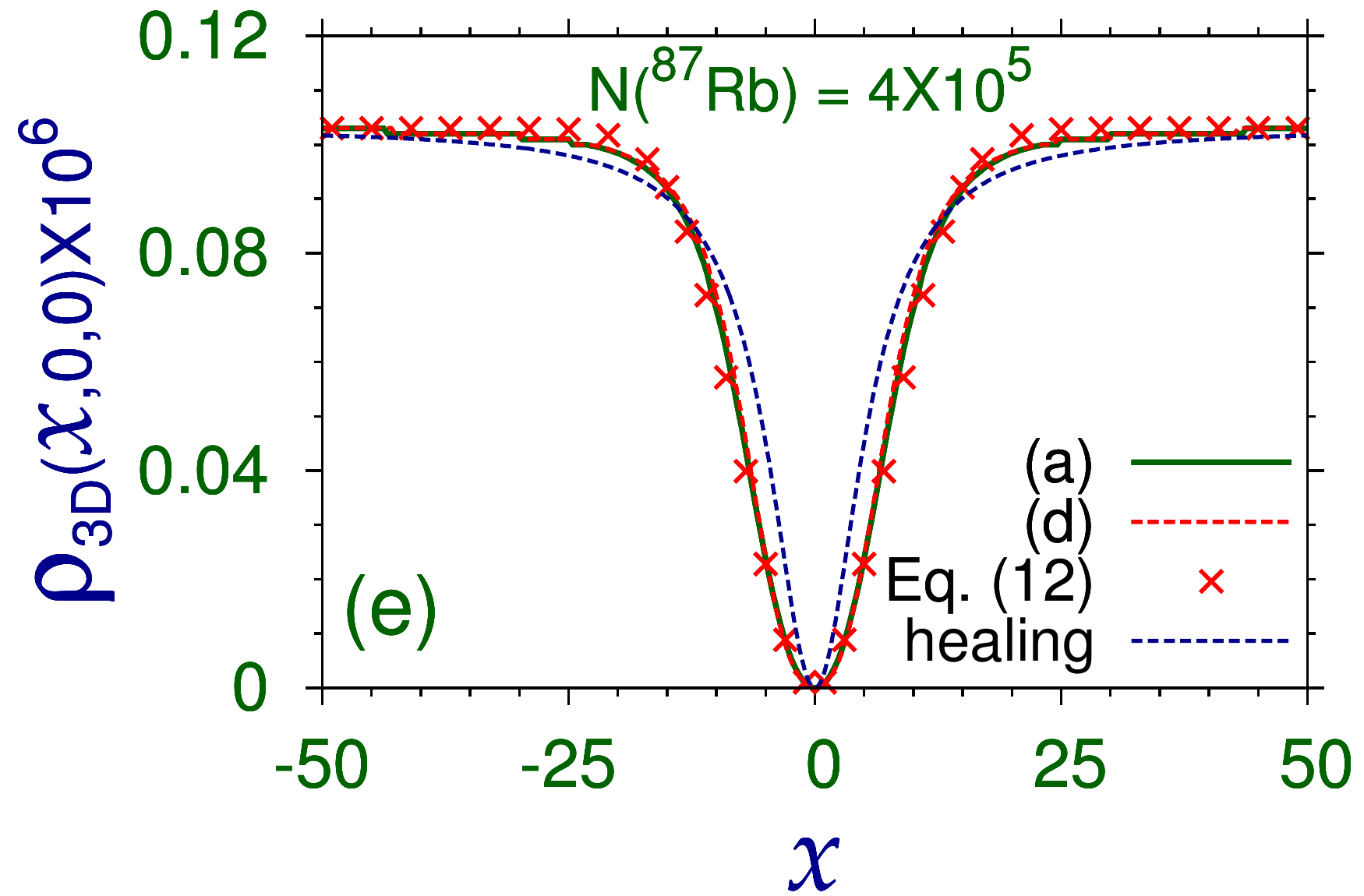}
\includegraphics[width=.49\linewidth,clip]{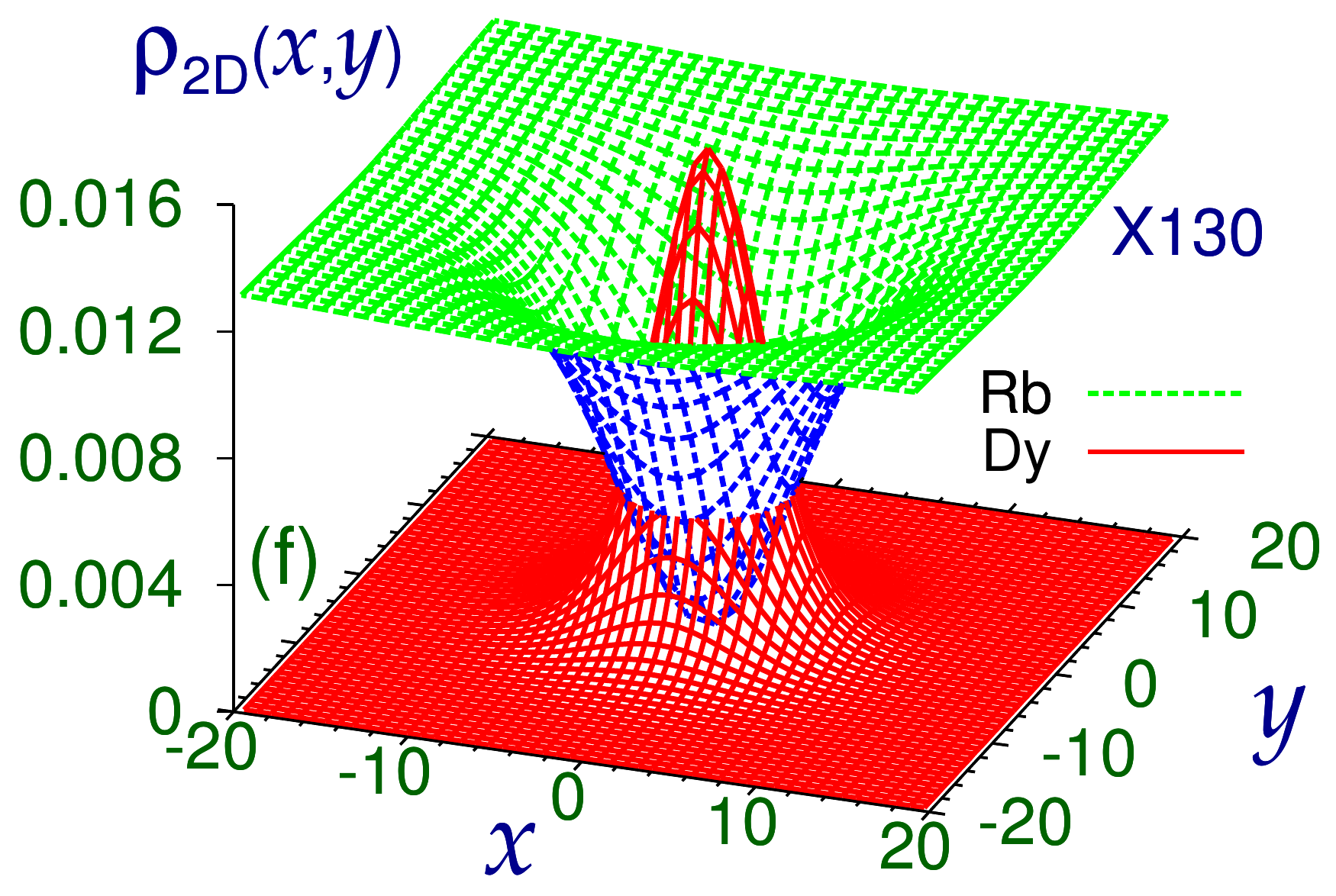}

\caption{ (Color online) (a) Integrated  1D density   $\rho_{1D}(x)$ and $\rho_{1D}(z)$ of a matter-wave soliton of
1000  $^{85}$Rb atoms in the   vortex core of a uniform $^{87}$Rb BEC of $4\times 10^5 $
atoms in a box of volume $100^3$ $\mu$m$^3$. The same of (b) 500, (c) 1000,
and (d) 2000  $^{164}$Dy atoms. (e)  3D density $\rho_{3D}(x,0,0)$ 
of the vortex core for (a) and (d) together with analytic fit (\ref{phi1}) and healing length 
estimate (healing) $\rho(x,0,0)= 0.00000103 x^2/(x^2+2\xi^2)$ with healing length $\xi\approx 4.072$ $\mu$m.
(f) Integrated 2D  density $\rho_{2D}(x,y)$ of the $^{87}$Rb vortex and the $^{164}$Dy soliton of 
(d). 
The  parameters  are  $a(^{87}$Rb-$^{85}$Rb) = $a$(Rb-Dy) =120$a_0$. 
Variables in all figures    are dimensionless.
}\label{fig1} \end{center}

\end{figure}

\section{Numerical Results in three dimensions}

\label{III}

In the 3D simulation of the binary vortex-soliton
 we consider the nondipolar $^{87}$Rb-$^{85}$Rb and the dipolar
$^{87}$Rb-$^{164}$Dy mixtures. The  $^{164}$Dy atom has the  magnetic moment  $\mu=10\mu_B$  \cite{ExpDy} with 
$\mu_B$ the Bohr magneton so that  the dipolar length $a_{\mathrm {dd}}(^{164}$Dy$) \approx 132.7a_0$ with $a_0$ the Bohr radius. 
We use scattering lengths $a(^{87}$Rb)= $a(^{87}$Rb-$^{85}$Rb) = 
 $a(^{87}$Rb-$^{164}$Dy) = $120a_0$  and take  $a(^{85}$Rb) and $a(^{164}$Dy) as variables. 
The experimental values of these scattering lengths are not known precisely. The exact values of the inter-species scattering lengths are not important for our analysis. These positive scattering lengths are used as they simulate the inter-species repulsion required for the formation of binary vortex-soliton.  Furthermore, if needed,
the variation of the scattering lengths 
can be achieved by the Feshbach resonance technique \cite{fesh}. 
We solve   (\ref{eq4}) and (\ref{eq5})  
by the split-time-step 
Crank-Nicolson method using both real- and imaginary-time propagations
  in Cartesian coordinates  
using a space step of  $ 0.2\sim 0.4$
and a time step of  $ 0.0025 \sim 0.005$ \cite{CPC}.  The dipolar  term is treated by a Fourier transformation  in momentum space using a convolution theorem  \cite{Santos01}. 
In all cases we take the length scale  
 $l =1$ $\mu$m and time scale $t_0= 2m(^{87}$Rb)$l^2 /\hbar =2.74$ ms.
The numerical simulation is performed in a cubic box, limited by  $x=y=z=\pm50$, containing 400000 $^{87}$Rb atoms of density $4\times 10^{11}$ /cc.

We consider a matter-wave soliton of 1000 nondipolar 
$^{85}$Rb atoms in the $^{87}$Rb vortex core and perform imaginary-time 
simulation.  A large intra-species attraction  with scattering length $a(^{85}$Rb) $=-80a_0$ was necessary to obtain a compact  soliton of small size. 
In the nondipolar case, unlike in 1D equation (\ref{model}),   
no 3D vortex-soliton can be obtained  for repulsive inter- and intra-species interactions.
An attractive intra-species interaction facilitates the formation of the   soliton.
For smaller values of  intra-species attraction, the size of the   soliton was larger resulting in computational difficulty. The 1D densities of the nondipolar soliton, defined by $\rho_{1D}(x)=\int dy\int dz |\phi({\bf r})|^2$, etc.,   are  plotted in figure \ref{fig1}(a).

 In the dipolar case, we consider     solitons of 500, 1000, and 2000 $^{164}$Dy atoms, in the $^{87}$Rb vortex core, with intra-species scattering lengths $a(^{164}$Dy) $=-40a_0, 10a_0$, and 
$30a_0$, respectively. Again the exact values of the scattering lengths are not important. These three values of scattering lengths simulate three distinct interactions: attractive, weakly repulsive and moderately repulsive. 
If needed, the intra-species and inter-species interactions can be  manipulated by independent 
optical and magmetic Feshbach resonances in a laboratory \cite{fesh}.   
The corresponding 1D densities of these   solitons are presented  in Figs. 
\ref{fig1} (b),  (c), and (d), respectively. The 3D density $\rho_{3D}({\bf r})=|\Phi_1({\bf r},t)|^2$ of the axially-symmetric vortex core is plotted 
in figure \ref{fig1}(e). We also show the  variational healing-length estimate of this density for an isolated vortex without the soliton: $\rho(x,y,0)= \rho_0 (x^2+y^2)/(\xi^2+x^2+y^2)$, where $\rho_0$ is the density away from the vortex core, and the healing length $\xi =1/\sqrt{8\pi \rho_0 a} \approx 4.072$ $\mu$m.  
The numerical solution of the GP equation for the isolated vortex agrees well with healing-length estimate. 
The deviation of the vortex core of the binary vortex-soliton from the healing-length estimate of an isolated 
vortex is due to the presence of the soliton which increases the size of the vortex core by inter-species repulsion. 
 In figure \ref{fig1}(f)  the integrated 2D densities $\rho_{2D}(x,y)\equiv \int dz |\phi
({\bf r})|^2$ for the vortex and the soliton corresponding to figure \ref{fig1}(d) are plotted.  
Figures \ref{fig1}(a) $-$ (f) show that the soliton 
is localized in the vortex core given by the  minimum of the vortex density plotted in figure    \ref{fig1}(e).

\begin{figure}[!t]

\begin{center}
\includegraphics[width=.48\linewidth,clip]{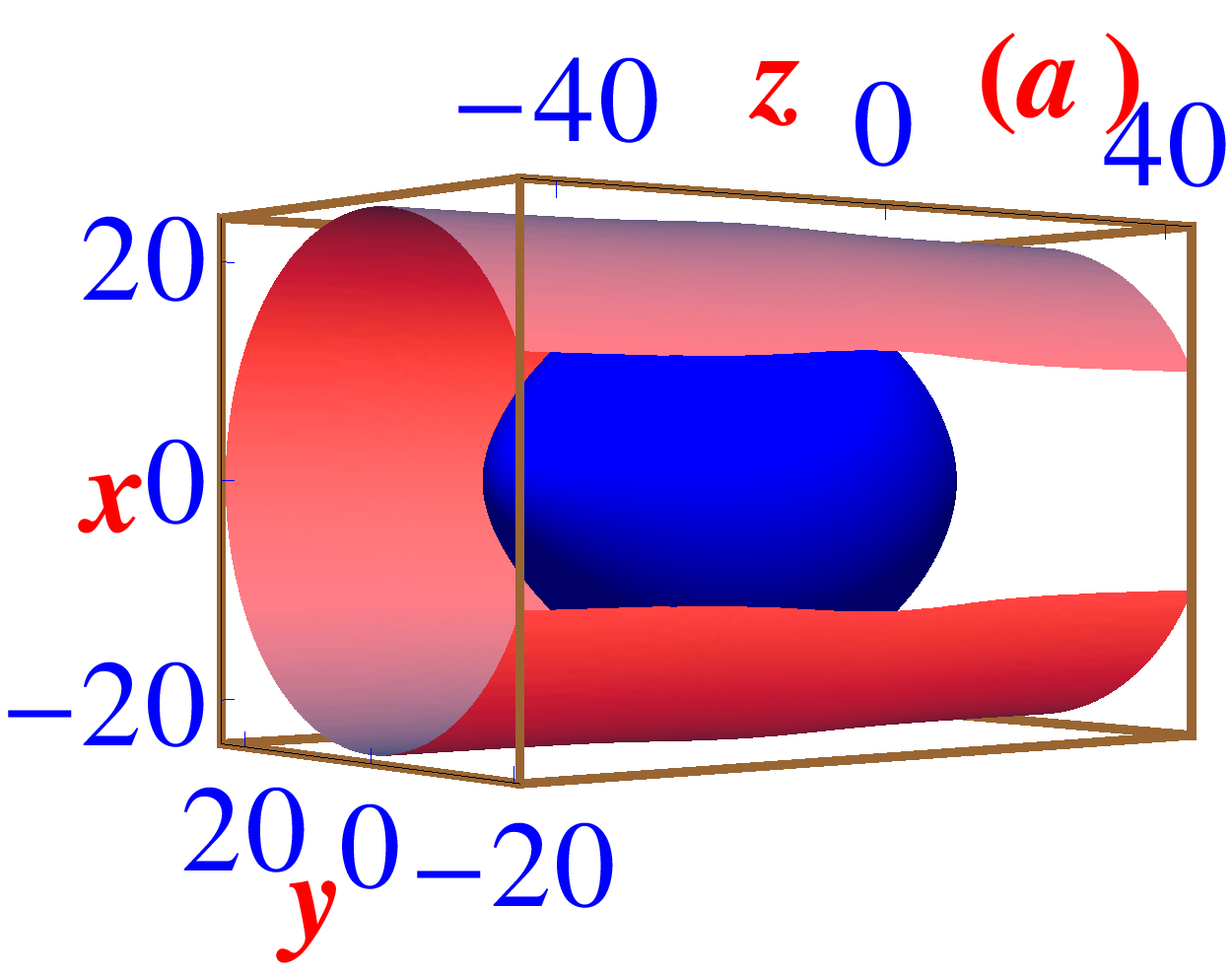}
 \includegraphics[width=.48\linewidth,clip]{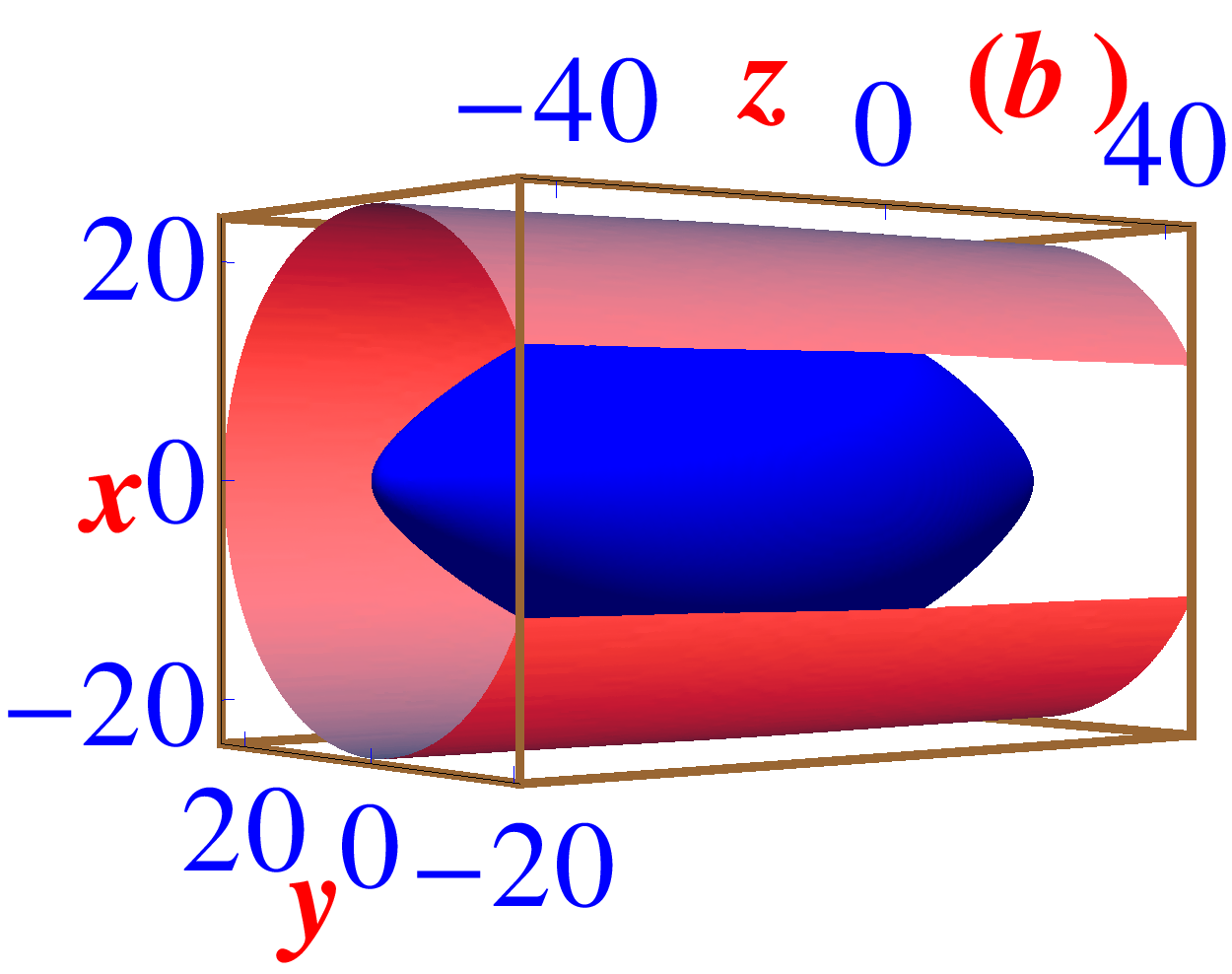}
\includegraphics[width=.48\linewidth,clip]{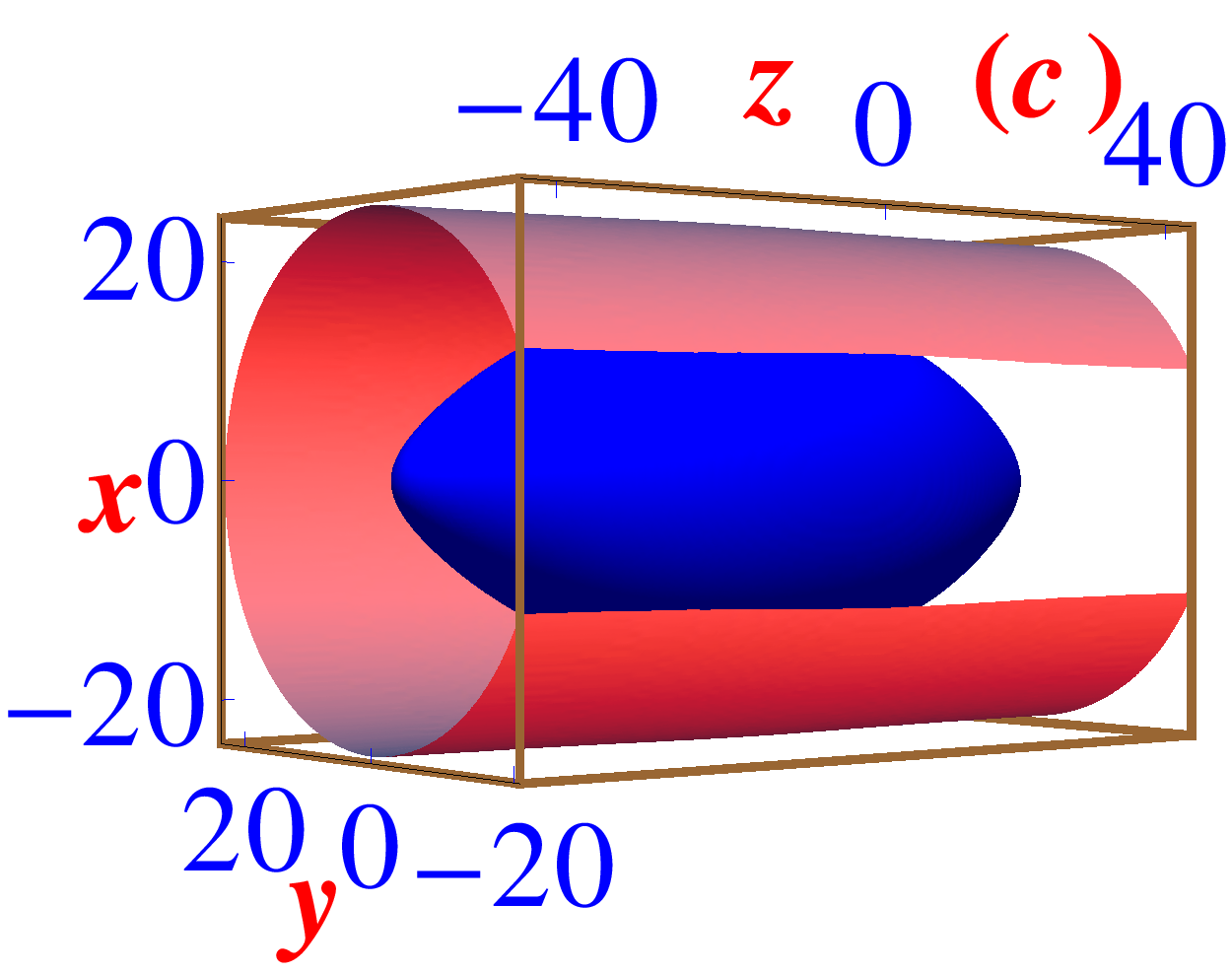}
 \includegraphics[width=.48\linewidth,clip]{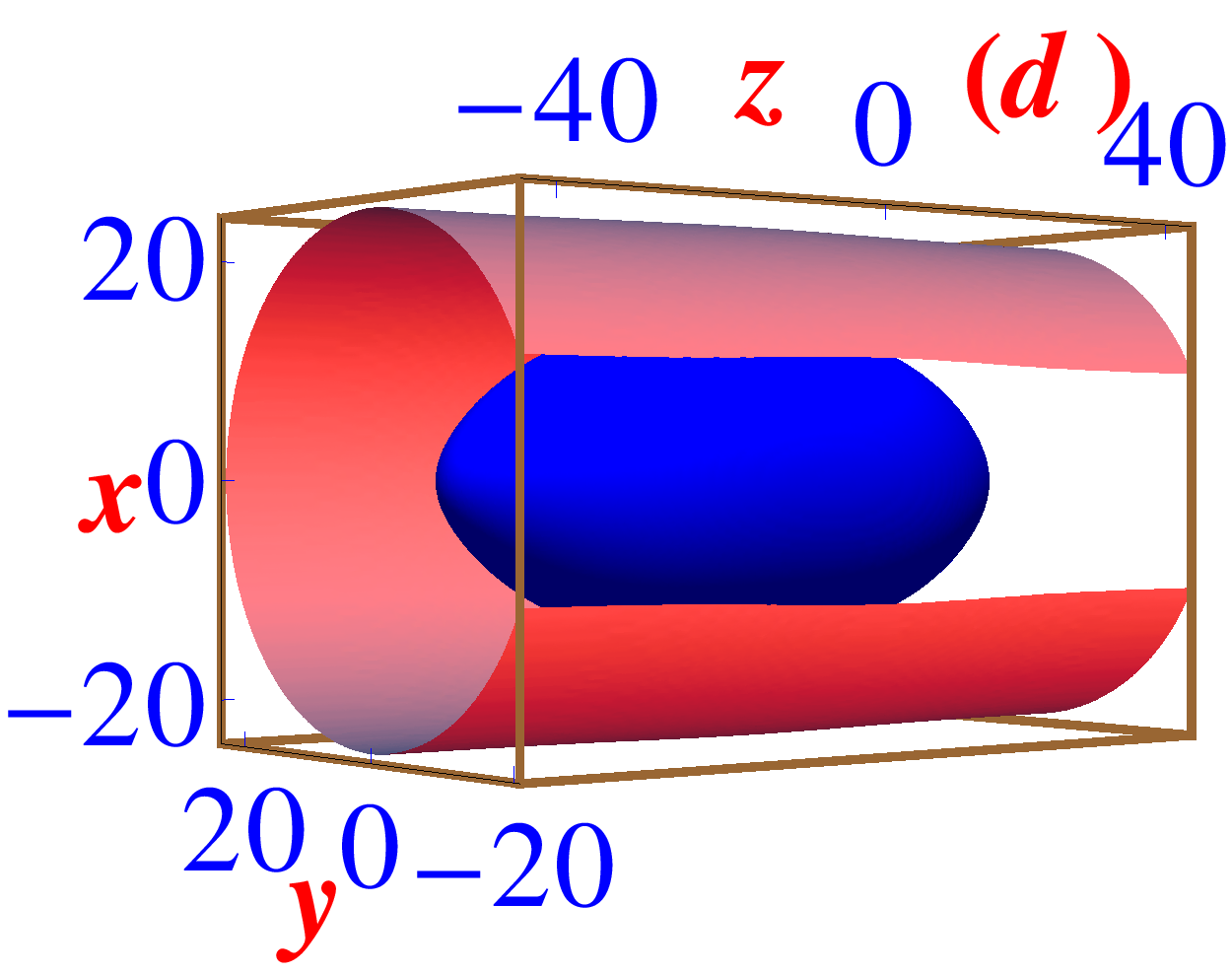}

\caption{(Color online)   3D isodensity contour of the binary  vortex-soliton, 
showing the vortex core (grey, pink in color) and the soliton (black, blue in color) profiles,   corresponding to (a) figure \ref{fig1}(a),  (b) figure \ref{fig1}(b), 
(c) figure \ref{fig1}(c), and  (d) figure \ref{fig1}(d).  Density on the contour is $ 10^{6}$ /cc.
}\label{fig2} 

\end{center}

\end{figure} 

A qualitative understanding of the formation of the matter-wave soliton can be obtained if 
we note that the densities  corresponding to  the vortex wave functions as presented in figure \ref{fig1}(e)  for the four cases studied above 
are practically the same  given by the following   function   
\begin{eqnarray}\label{phi1}
\Phi_1^2({\bf r})=A\big(1 - e^{-(x^2+y^2) \delta}\big)
\big(1+\nu -\nu e^{-z^2\delta }\big),
\end{eqnarray}
with 
$ A= 0.00000103, \delta = 0.01,  \nu =0.25.$ The function (\ref{phi1}) with $\nu=0$ simulates a vortex in a uniform BEC. A non-zero value of $\nu$ includes the effect of a small distortion of the vortex 
inside the soliton 
due to the presence of the soliton. 
The analytic density (\ref{phi1})
is also plotted in figure \ref{fig1}(e) in good agreement with the numerical result.
We perform an approximate variational calculation 
for the formation of a
soliton with vortex density (\ref{phi1}) substituted in  (\ref{eq4}) using the  following Gaussian 
trial  function for the soliton:
\begin{equation}
\phi_2= \frac{\pi^{-3/4}}{w_\rho\sqrt{w_z}} \exp\Big[
-\frac{x^2+y^2}{2w_\rho^2}-\frac{z^2}{2w_z^2}\Big],
\end{equation}
where $w_\rho$ and $w_z$ are the widths.
The Lagrangian  (\ref{eq4})  can be written as 
\begin{eqnarray}
L &=\frac{1}{2}\int d{\bf r} \Big[m_{12}N_2 |\nabla \phi_2|^2+ N_2 g_2 \phi_2^4+2N_2g_{21}|\Phi_1|^2
|\phi_2|^2\nonumber \\
& +\int  d{\bf r}\int d{\bf r'} {N_2}g_{{\mathrm {dd}}}V_{{\mathrm {dd}}}({\bf R})|\phi_2({\bf r})|^2
|\phi_2({\bf r'})|^2\Big],\\
&=\frac{N_2m_{12}}{2}\Big[ \frac{1}{w_\rho^2}+ \frac{1}{2w_z^2}\Big]+\frac{N_2^2m_{12}[a_2-a_{{\mathrm {dd}}}f(\kappa)]}{\sqrt{2\pi}w_\rho^2w_z}\label{lag}
\nonumber \\ 
 &+ \frac{N_2 g_{21}A\delta w_\rho^2}
{ e^2(w_\rho)}
\Big[ 1+\nu -\frac{\nu}{e( w_z)}\Big]
\end{eqnarray}
where $\kappa=w_\rho/w_z$,
$f(\kappa)=[1+2\kappa^2-3\kappa^2 d(\kappa)]/(1-\kappa^2), d(\kappa)=\arctan (\sqrt{\kappa^2-1})/\sqrt{\kappa^2-1}$,  $e(x)=\sqrt{1+\delta x^2}$.
Minimizing  Lagrangian   (\ref{lag})
we find the following conditions to determine the widths of the soliton \cite{cr,1D}
\begin{eqnarray}\label{a1}
&\frac{1}{w_\rho^3}+\frac{N_2e(\kappa)}{\sqrt{2\pi}w_\rho^3w_z}-\frac{2w_\rho g_{21}A \delta}{e^4(w_\rho)m_{12}}\Big[1+\nu-\frac{\nu}{e(w_z)}  \Big]=0, \\
&\frac{m_{12}}{w_z^3}+\frac{2N_2m_{12}h(\kappa)}{\sqrt{2\pi}w_\rho^2w_z^2}
-\frac{2g_{21}A\nu \delta^2 w_\rho^2 w_z}{e^2(w_\rho)e^3(w_z)}=0,
\label{a2}
\end{eqnarray}
where   $e(\kappa)=2a_2-a_{\mathrm{dd}}
[2-7\kappa^2-4\kappa^4+9\kappa^4 d(\kappa)]/(1-\kappa^2)^2, h(\kappa)=a_2-a_{\mathrm{dd}}
[1+10\kappa^2-2\kappa^4-9\kappa^2 d(\kappa)]/(1-\kappa^2)^2.$  
A solution of  (\ref{a1}) and (\ref{a2}) determines the widths, and hence the sizes
of the soliton. For the solitons of figure \ref{fig1} (a) $-$ (d) the
 numerical and analytic root-mean-square (RMS) sizes  are  given in 
table \ref{table1}. The analytic result presented is not a variational solution of the full dynamics given
by  (\ref{eq5}) and (\ref{eq4}) and hence there are no variational bounds on the energy or sizes. 
Nevertheless, the approximate analytic results of table \ref{table1}, in reasonable agreement with the numerical results,  provide a qualitative understanding of the formation of the soliton.

\begin{table}[b]
\caption{Numerical and analytic RMS sizes of the four solitons presented 
in Figs. \ref{fig2}(a)$-$(d).}
\label{table1}

\begin{tabular}{ c c c c c c}
 \hline 
   & $\langle x,y\rangle_{num}$&  $\langle x,y\rangle_{anal}$&
  $\langle z\rangle_{num}$&  $\langle z\rangle_{anal}$ & $\langle z/ x\rangle_{num}$\\
 \hline
(a)& 4.89  & 4.762 &  6.97 & 7.519&  1.43\\
(b) & 4.66  &  4.341 & 11.05 & 10.557 &2.37\\
(c) & 4.47 &  4.226 & 10.90 & 11.711 &2.44\\
(d) & 3.79 &  3.870 & 9.72 & 11.884 &2.56\\
 \hline
\end{tabular}
\end{table}

The density for  the dipolar matter-wave  solitons is strongly  anisotropic with distinct 1D densities $\rho_{1D}(x)$ and $\rho_{1D}(z)$ as can be found in Figs. \ref{fig1}(b)$-$(d), whereas in the nondipolar case, shown in figure \ref{fig1}(a), these two densities are nearly equal. 
The anisotropy in the shape of the soliton arises partly due to the anisotropy of the dipolar interaction and partly due to the anisotropy of the vortex core. 
 The increase in total dipolar interaction makes the soliton more elongated (prolate) along the polarization $z$ direction. This is reflected 
in the ratio of numerical RMS sizes $\langle z \rangle/\langle x \rangle$ shown in table \ref{table1}. 

The 3D isodensity contours of the four matter-wave solitons of Figs. \ref{fig1}(a)$-$(d) together with the respective vortex cores 
are shown in Figs. \ref{fig2} (a)$-$(d), respectively. The central prolate spheroid
is the soliton and the outer shell is the vortex core.  
The net dipolar interaction and hence the anisotropy of the soliton  increases with the number of dipolar  $^{164}$Dy atoms. The 3D profile of the soliton is   more prolate, in Figs. \ref{fig2}(b)$-$(d),
due to dipolar interaction, compared to that in figure \ref{fig2}(a) for the nondipolar case. The  nondipolar soliton is also slightly prolate due to the axially-symmetric inter-species repulsion of the vortex.

\begin{figure}[!t]

\begin{center}
\includegraphics[width=.49\linewidth,clip]{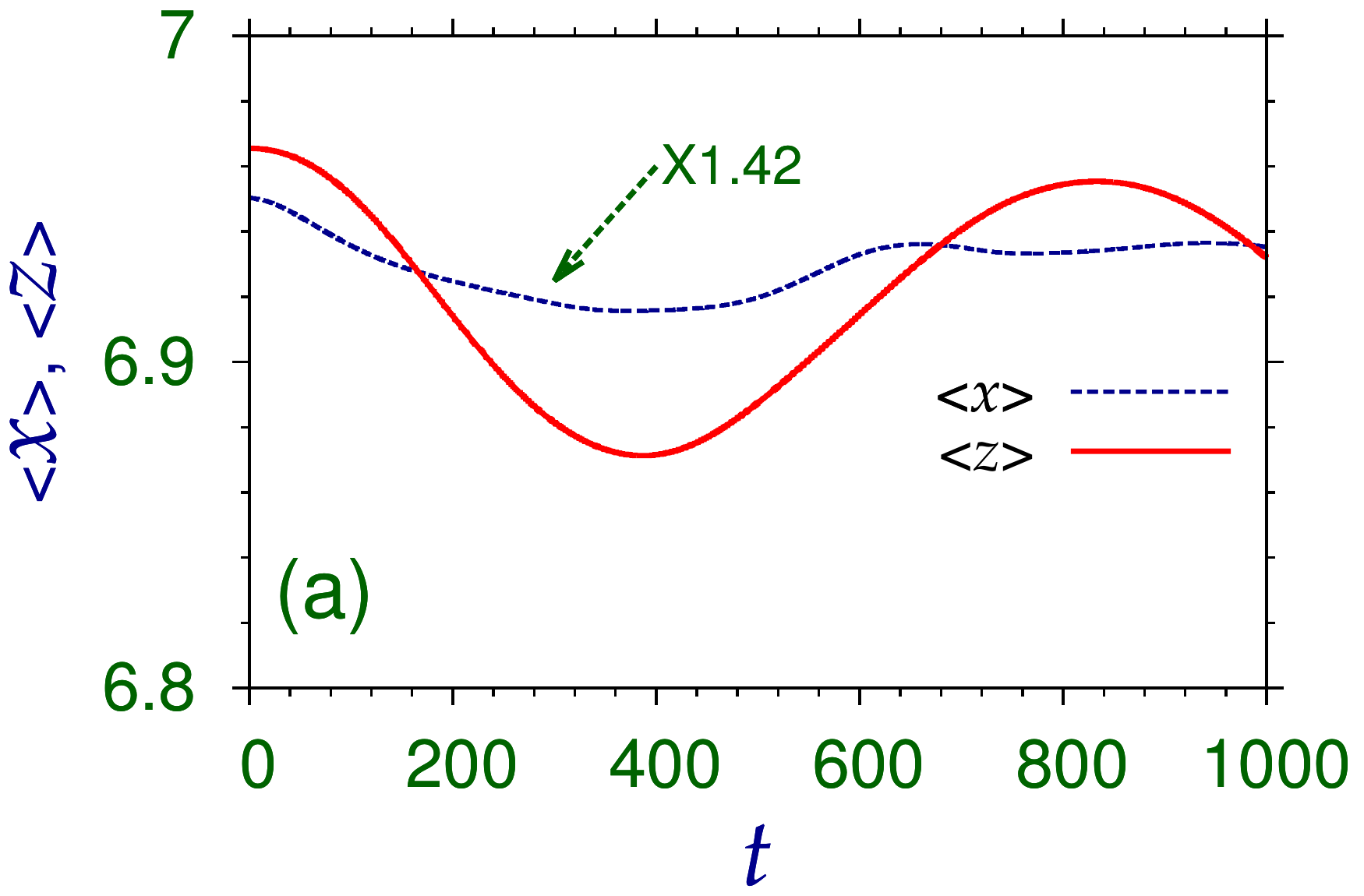}
  \includegraphics[width=.49\linewidth,clip]{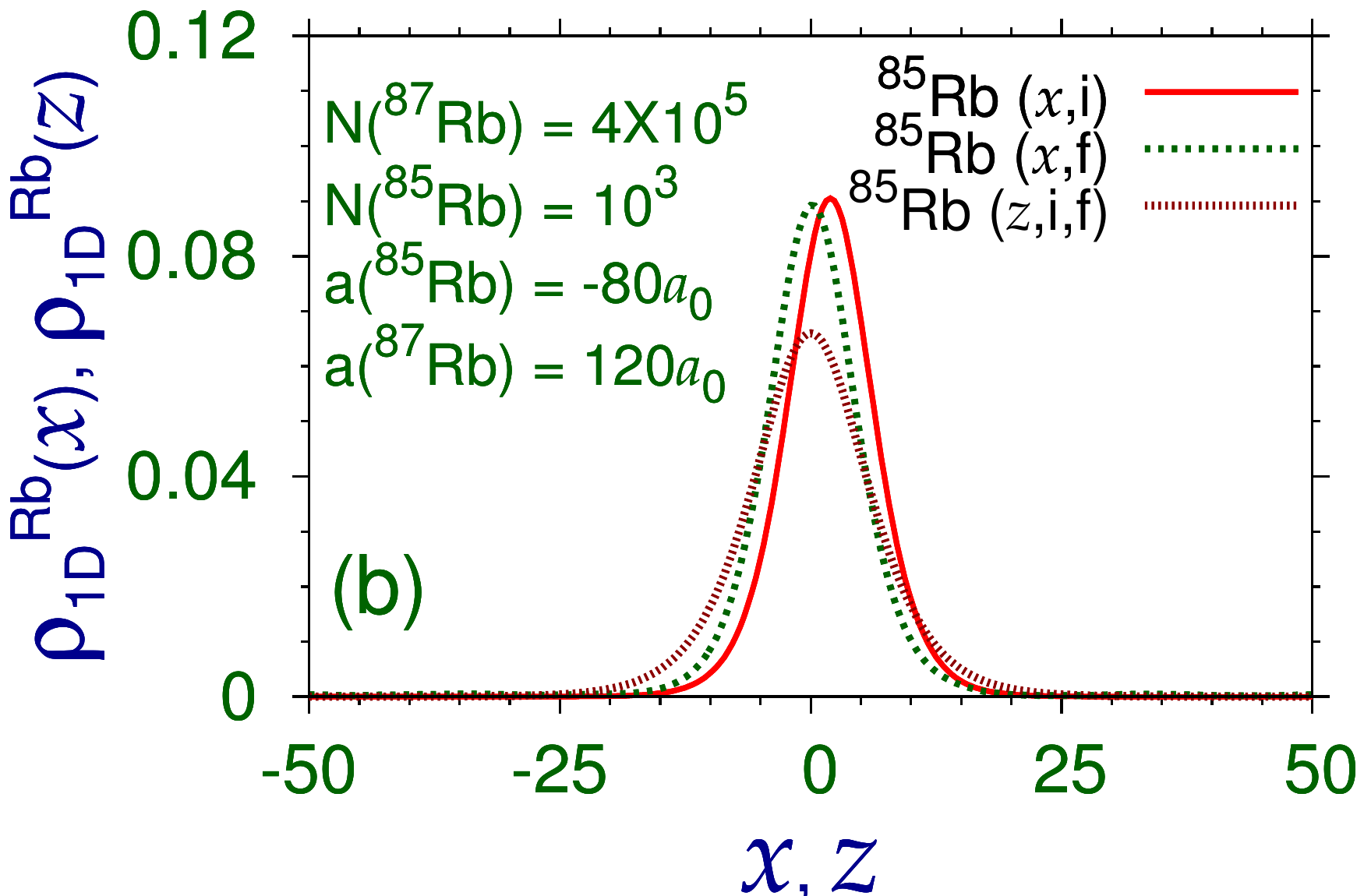}

\caption{(Color online) (a) RMS sizes $\langle x\rangle, \langle z\rangle$ during breathing oscillation of the nondipolar soliton of Figs. \ref{fig1}(a) and  \ref{fig2}(a) initiated by a sudden change of $a(^{85}$Rb) from $-80a_0$ to $-80.5a_0$. (b) Initial (i) and final (f)  1D densities along $x$ and $z$ directions after real-time dynamics of the  vortex-soliton of figure \ref{fig2} (a) during 1000 units of time. The dynamics is started 
after shifting the initial soliton  along $x$ direction through 2 units of length and maintaining the initial vortex position unchanged.  In both (a) and (b) the initial 
state was the imaginary-time stationary profile of the vortex-soliton of figure 
   \ref{fig2}(a).
}\label{fig3} 

\end{center}

\end{figure}

 \begin{figure}[!t]

\begin{center}
\includegraphics[width=.53\linewidth]{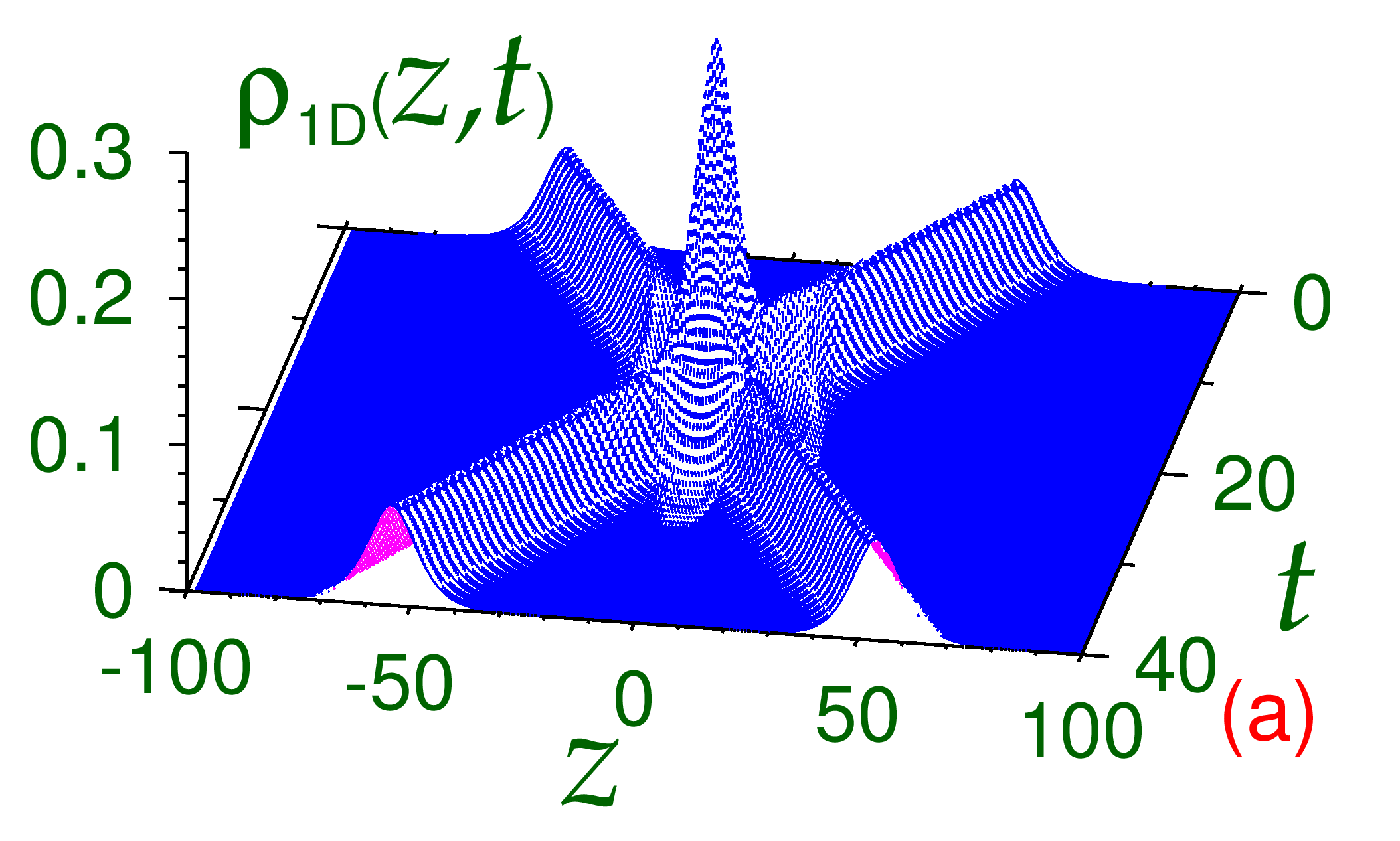}
 \includegraphics[width=.41\linewidth]{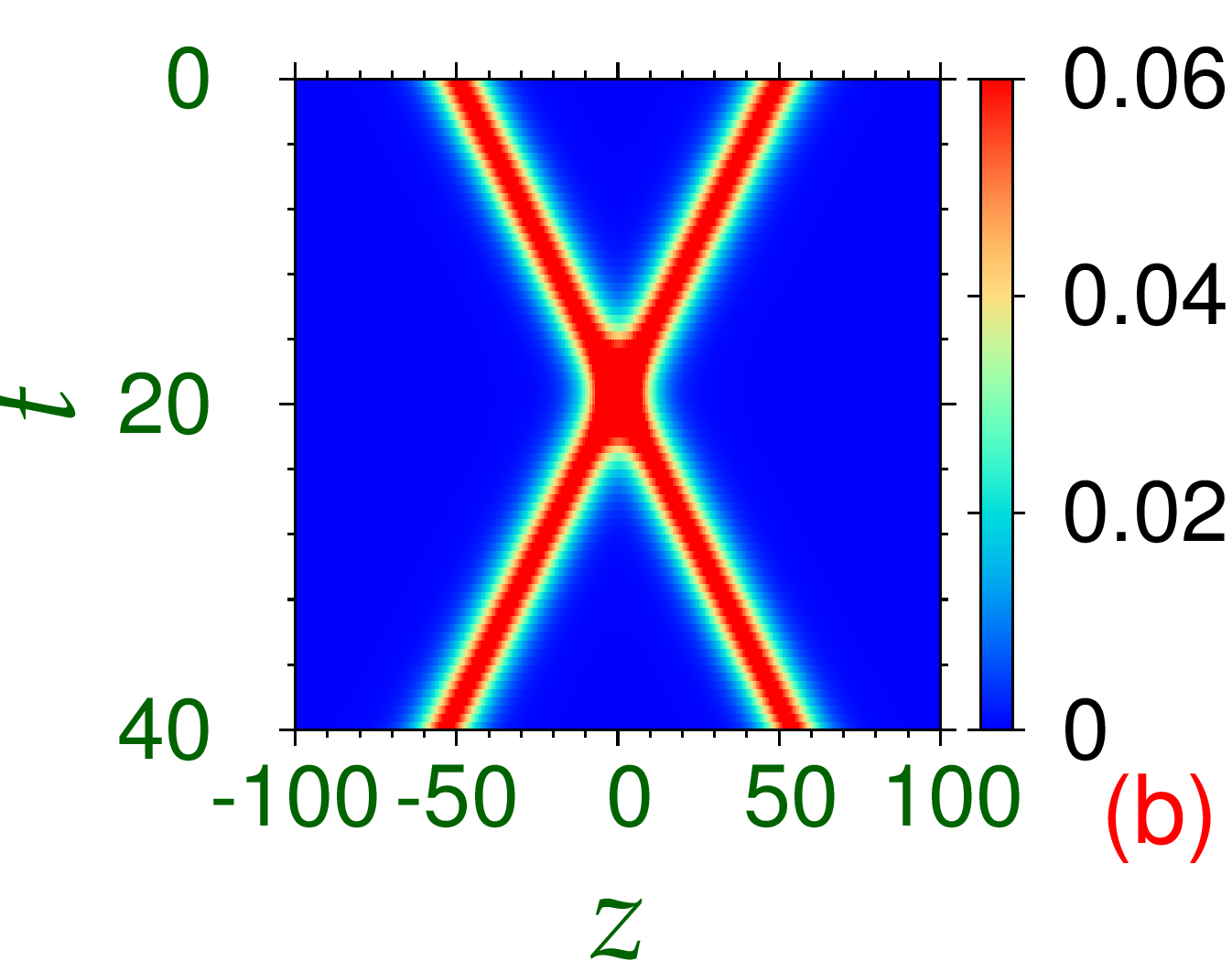}
\includegraphics[width=.53\linewidth]{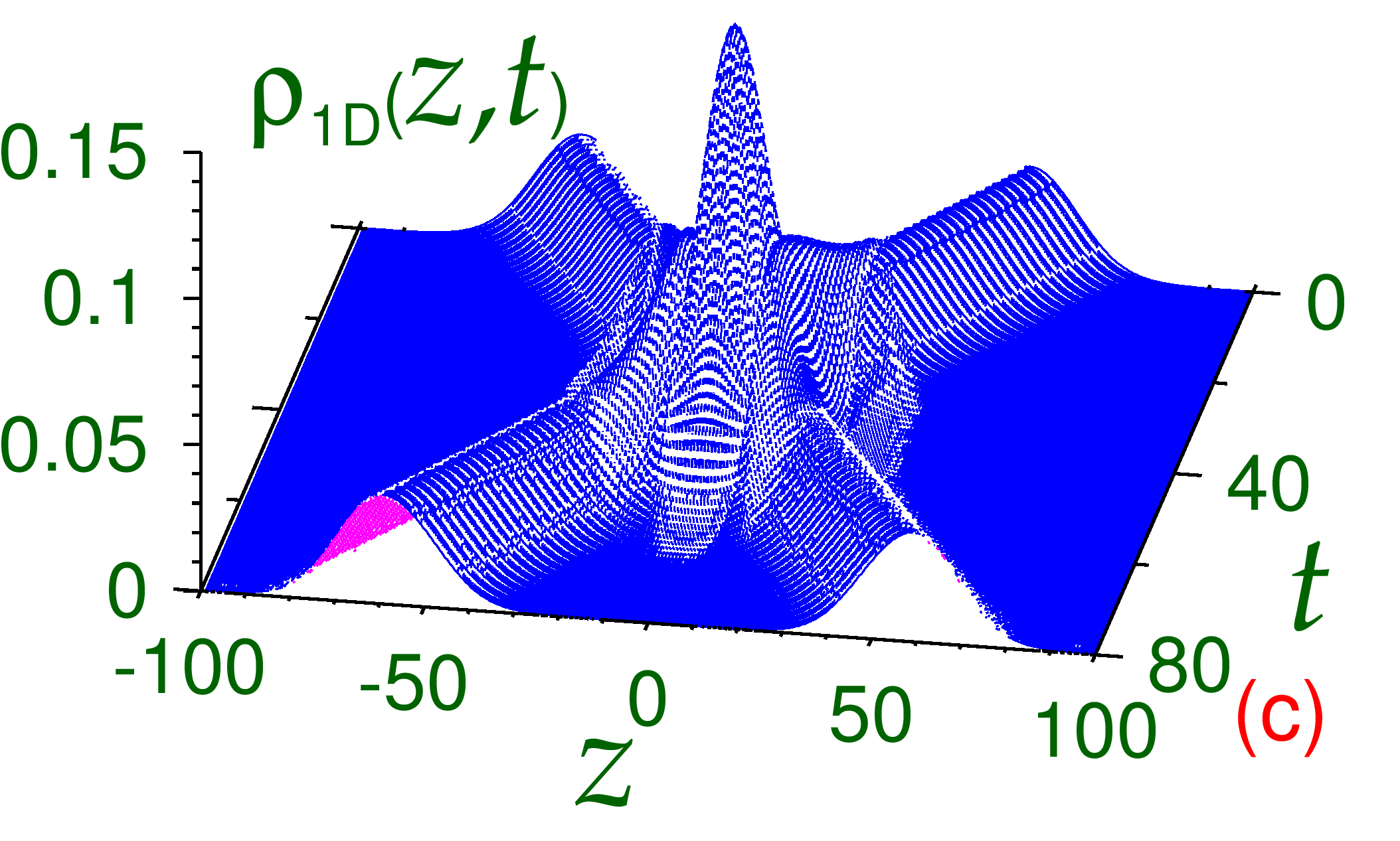}
 \includegraphics[width=.41\linewidth]{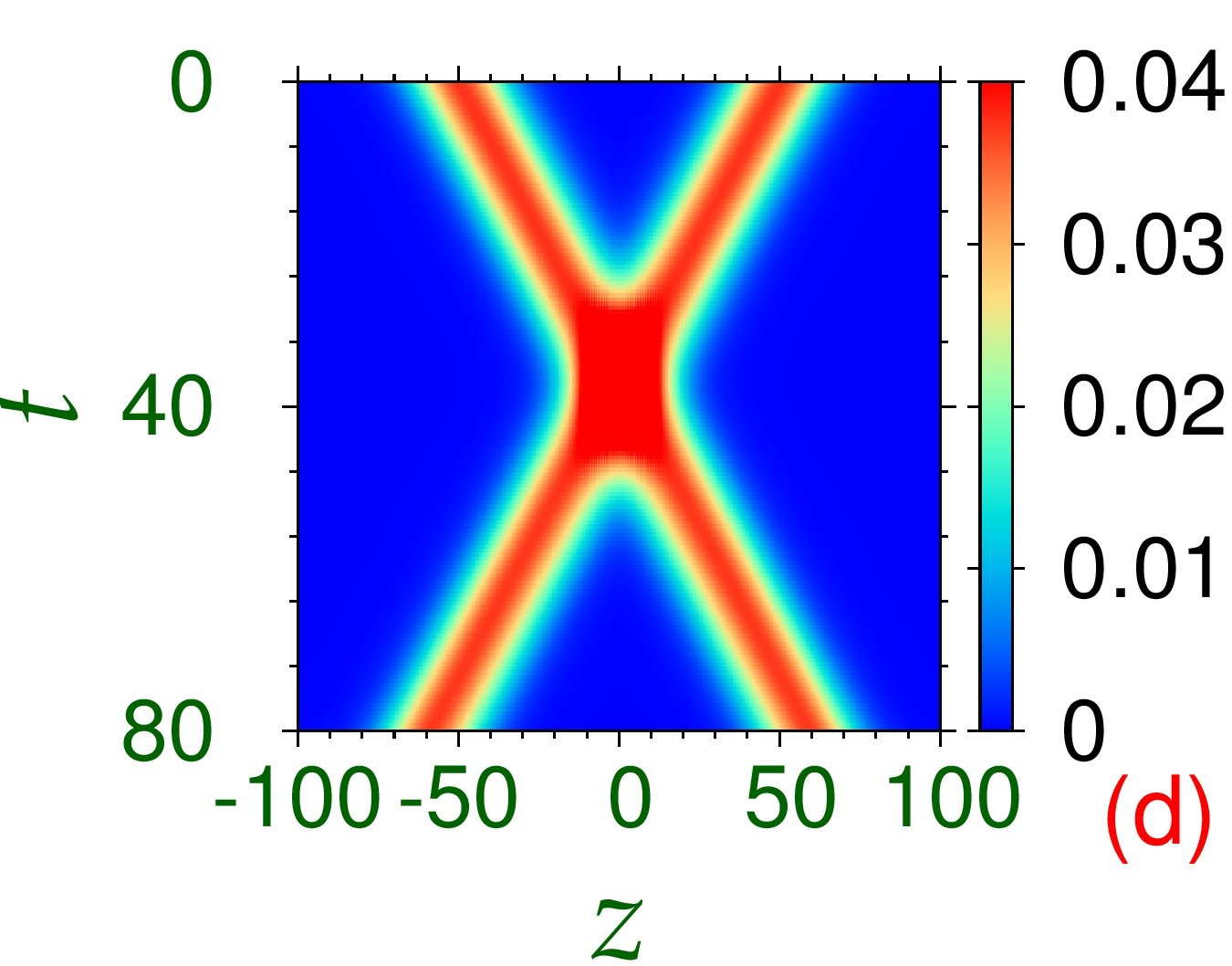}

\caption{ (Color online)
(a) The 1D density   $\rho_{1D}(z,t)$  and (b) its contour plot during  
 collision of two nondipolar  solitons of 1000 $^{85}$Rb atoms
of figure \ref{fig2} (a) 
initially placed at $z=\pm 50 $, upon real-time
propagation. The initial wave functions are multiplied by  $\exp(\pm i 20 z)$ to set them in motion.   
The same for two dipolar  solitons of 1000 $^{164}$Dy atoms
 of figure \ref{fig2} (c) are shown in (c)  and (d).    
}\label{fig4} \end{center}

\end{figure}

By carefully adjusting the parameters of the mean-field model equations a balance between attraction and repulsion can be achieved to obtain a vortex-soliton 
similar to a Townes soliton in 2D \cite{townes}, which is weakly unstable and collapses upon a small perturbation \cite{sptem1}. So it is quite relevant to establish the stability of the vortex-soliton.
To demonstrate  the stability of the soliton, we consider the one in figure \ref{fig2}(a)
and 
 subject the stationary state(s) obtained by  imaginary-time 
propagation to real-time propagation introducing a small perturbation, e.g., jumping intra-species scattering length 
$a(^{85}$Rb) from $-80a_0$ to $-80.5a_0$ at $t=0$. Long-time stable oscillation  of the resultant RMS  sizes, 
illustrated  in figure \ref{fig3} (a), guarantees the stability of the soliton. To test the stability of the soliton after a small displacement along $x$ direction, we performed real-time simulation of the   vortex-soliton of figure \ref{fig2} (a) after displacing it through 2 units of length along $x$ direction. The soliton is found to come back to its stable position on the $z$ axis at the center of the vortex core without executing oscillatory motion along $x$ axis. In figure \ref{fig3} (b) the plot of initial and final 1D densities of the soliton after 1000 units of time confirms the stability. This transverse stability of the vortex-soliton is important for an experimental realization.


Next we study the head-on collision between  two solitons moving along the vortex core in opposite directions. The imaginary-time profile of the vortex-soliton   of figure \ref{fig2}(a)  is used as the initial function in the real-time simulation of collision, with two identical solitons placed at $z=\pm 50. $  To set the solitons in motion along the $z$ axis in opposite directions the soliton wave functions are multiplied by $\exp(\pm i v z), v=20$. To illustrate the dynamics upon real-time simulation, we plot  the time evolution of  1D density $\rho_{1D}(z,t)$ in figure \ref{fig4}(a) and its contour plot in figure \ref{fig4}(b). The same for the   collision of two solitons of figure \ref{fig2}(c) are shown in  Figs.  \ref{fig4}(c) and (d).
In Figs. \ref{fig4}(a) and (b), the dimensionless velocity of a soliton  is $\sim 2.5$, whereas in Figs.  \ref{fig4}(c) and (d) this velocity is $\sim 1.25$.
 The quasi-elastic nature of collision is established at a relative velocity (two times the velocity of a single soliton)
 of about $2\sim 5$
 in dimensionless units.





\section{Summary}

\label{IV}

Summarizing, we demonstrated the possibility of the creation of a nondipolar or dipolar
matter-wave soliton in the vortex core of a uniform nondipolar
BEC. The soliton is localized by a strong inter-species 
repulsion. 
 This binary vortex-soliton is a stable stationary state. A dipolar soliton can be created for 
repulsive inter- and intra-species contact interactions. However,  for the creation of a nondipolar 
soliton an attractive    intra-species contact interaction in the soliton is necessary.
The 
matter-wave soliton can move with a constant velocity  along the vortex core without any deformation.   
The stability of the vortex-soliton is demonstrated by a stable oscillation of the soliton upon 
a small perturbation in real-time simulation using the initial states obtained in imaginary-time propagation, viz. figure \ref{fig3}. 
At medium velocities, the collision between the two solitons is   quasi elastic with 
no visible deformation, viz. figure \ref{fig4}. 

The techniques of generating a vortex
in BECs \cite{vorbec}
are  well known, hence   a  binary vortex-soliton can be realized in experiments.
 Here, for the sake of computational simplicity, we considered the vortex in a uniform BEC.  However, for an experimental observation,  a small matter-wave soliton can be created in the   core of a large low-density weakly-trapped BEC vortex. 
To  achieve this 
the binary vortex-soliton should be realized under a weak harmonic trap on both components in a laboratory 
and eventually the weak trap on the soliton  should be slowly removed to obtain the trapless soliton 
in the vortex core maintaining the weak trap on the vortex. The weak initial harmonic trap on both components will 
bring the soliton inside the vortex core. We tested the stability of the binary vortex-soliton under a small transverse perturbation. The transverse confining  force on the soliton due to inter-species repulsion brings the 
soliton back to the center inside the vortex core. An interesting future work would be to study the possibility of creating a binary vortex-soliton in two components of a spin-orbit coupled BEC with one component supporting a coreless vortex \cite{core} and the other supporting a soliton.


\ack
We thank the Funda\c c\~ao de Amparo 
\`a
Pesquisa do Estado de S\~ao Paulo (Brazil)  and  the
Conselho Nacional de Desenvolvimento   Cient\'ifico e Tecnol\'ogico (Brazil) for partial support.


\section*{References}
\begin{thebibliography}{99}

\bibitem{book} Kivshar Y S and  Agrawal G 2003 {\it Optical Solitons: From Fibers to 
Photonic Crystals} (San Diego, Academic Press)



\bibitem{rmp} Kivshar Y S and  Malomed B A 1989 \RMP
  {\bf 61} 763 

 Abdullaev F K,  Gammal A,  Kamchatnov A M  and 
  Tomio L 2005 
{\it Int. J. Mod. Phys.} B {\bf 19} 3415 

 
Perez-Garcia V M,  Michinel H and  Herrero H 1998 \PR A
 {\bf 57} 3837

 

\bibitem{1}  Strecker K E,  Partridge G B,  Truscott A G  and  Hulet R G 2002   Nature  {\bf 417} 150 

 Khaykovich L,  Schreck F,  Ferrari G,  Bourdel T,  Cubizolles J,  Carr L D, Castin Y  and  Salomon C
2002
 {\it Science} {\bf 256} 1290 


\bibitem{3}  Cornish S L,  Thompson S T  and  Wieman C E  2006 \PRL {\bf 96} 170401 (2006).


\bibitem{dark} Burger S,  Bongs K,  Dettmer S,  Ertmer W,  Sengstock K,  Sanpera A,  Shlyapnikov G V and  Lewenstein M 1999 {\it Phys. Rev. Lett.} {\bf 83} 5198 
\bibitem{dark1}
  Denschlag  J {\it et al.} 2000 {\it Science} {\bf 287} 97 

 



\bibitem{sptem1} Malomed B A,   Mihalache D,  Wise F 
and  Torner L 2005 {\it J. Opt. B} {\bf 7} R53 

 

\bibitem{4-5}
 Edmundson D E and  Enns R H  1992 {\it Opt. Lett.} {\bf 17} 586 

  Zhong W P,  Xie R H,  Beli\'c M,  Petrovic N  and  Chen  G 2008
\PR A {\bf 78} 023821 
%


\bibitem{3dvor} Mihalache D,  Mazilu D,  Crasovan L-C,  Towers I,  Buryak A V,  Malomed B A,  Torner L,  Torres  J P  and  Lederer F 2002
\PRL {\bf 88} 073902 

 


\bibitem{6-8}
 Kanashov A A and  Rubenchik A M  1981 {\it Physica} D {\bf 4} 122

 Malomed B A,  Drummond P,  He H,  Berntson A, 
 Anderson D and  Lisak M 1997 \PR E {\bf 56} 4725 

 Serkin V N and  Hasegawa A 2000 \PRL  {\bf 85} 4502



\bibitem{9} Adhikari S K  2004  {\it Phys. Rev.} A {\bf 69} 063613 

 Adhikari S K 2005  \PR E {\bf 71} 016611 
\bibitem{1dopttem}
 Mollenauer L F,  Stolen R H and   Gordon J P 1980 \PRL    {\bf 45}
1095 

\bibitem{1doptsp} Christodoulides D N and   Joseph R I 1988   {\it Opt. Lett.} {\bf 13} 794

Eisenberg H S,  Silberberg Y,  Morandotti R,   Boyd A R and   Aitchison J S 1998 \PRL  {\bf 81} 3383 



\bibitem{2dopt} Fleischer J W,  Segev M,  Efremidis N K  and   Christodoulides D N 2003  Nature 
{\bf 422} 147 

 Torruellas W E,  Wang Z,  Hagan D J,  VanStryland E W,  Stegeman G I,  Torner L and  Menyuk C R
1995 \PRL {\bf 74} 5036






\bibitem{3dopt} Minardi S {\it et al.} 2010  \PRL {\bf 105} 263901 







\bibitem{ExpDy} Lu M,  Burdick N Q,  Youn S H and  Lev B L 2011 {\it Phys.
Rev. Lett.} {\bf 107} 190401 


\bibitem{ExpEr} Aikawa K,  Frisch A,  Mark M,  Baier S,  Rietzler A,  Grimm R and Ferlaino F 
2012 {\it Phys. Rev. Lett.} {\bf 108} 210401 





\bibitem{cr}  Lahaye T,  Koch T,  Fr\"ohlich B,  Fattori M,  Metz J,  Griesmaier A,  Giovanazzi S and   Pfau T 2007  {\it Nature } {\bf 448} 672  

 Koch T,  Lahaye T,  Metz J,  Fröhlich B, Griesmaier A and   Pfau T 2008  Nature Phys. {\bf 4} 218 





\bibitem{1D} Young-S L E,  Muruganandam P  and   Adhikari S K 2011 \jpb {\bf 44} 101001 



 

 
 
 \bibitem{2D}
 Nath R,  Pedri P and  Santos L 2009
{\it Phys. Rev. Lett.} {\bf 102} 050401 

 Pedri P  and  Santos L 2005
{\it Phys. Rev. Lett.} {\bf 95} 200404  
 
 Tikhonenkov I I,  Malomed B A and  Vardi A 2008 {\it Phys. Rev. Lett.} {\bf 100} 090406 

 
 
\bibitem{ol2D}
 Adhikari  S K and   Muruganandam P 2012
 \jpb  
{\bf 45} 045301 
 
 Adhikari S K and   Muruganandam P 2012
 \PL A {\bf 376}  2200 

\bibitem{stablesol} Adhikari S K 2014   \PR A {\bf 89} 043615 
 
\bibitem{luca} Salasnich L,  Parola A  L. Reatto L 2002 \PR A  {\bf 65} 043614  


\bibitem{mfb2}
 Young-S L E and  Adhikari S K 2012 \PR A  {\bf  86} 063611 

 Young-S L E and  Adhikari S K  2013
\PR A {\bf 87}    013618 
 
 
 \bibitem{fesh}
 Inouye S,  Andrews M R,  Stenger J,  Miesner H-J,  Stamper-Kurn D M and    Ketterle W Nature 
1998  {\bf 392} 151 






%
 
\bibitem{CPC} 
 Kishor Kumar R,  Young-S. L E,  Vudragovic D,  Balaz A,  Muruganandam P and   Adhikari S K
2015  {\it  Comput. Phys.
Commun.}  {\bf 195}   117

 Muruganandam P  and  Adhikari S K 2009 {\it  Comput. Phys.
Commun.} {\bf 180} 1888 

  Vudragovic D,  Vidanovic I,
 Balaz A,  Muruganandam P and  Adhikari S K  2012 {\it Comput.
Phys. Commun.}  {\bf 183} 2021 




 Adhikari S K  and  Muruganandam P  2002 \jpb {\bf 35}
 2831 



\bibitem{Santos01}Goral  K and Santos L 2002 \PR A  {\bf 66} 023613 

 Yi S and  You L 2001  \PR A {\bf 63} 053607 



\bibitem{townes} Chiao R Y,  Garmire E and  Townes C H 1964
{\it Phys. Rev. Lett.} {\bf 13} 479 
 


\bibitem{vorbec} Fetter A   2009 {\it Rev. Mod. Phys.} {\bf 81} 647 
 

 
   Matthews M R,  Anderson B P,  Haljan P C,  Hall D S,  Wieman C E and  Cornell E A 1999 {\it Phys. Rev. Lett.} {\bf 83} 2498 


\bibitem{core} Wright K C,   Leslie L S,   Hansen A  and  Bigelow N P 2009  {\it Phys. Rev. Lett.}
{\bf 102} 030405

 Choi J-Y,  Kwon W J and  Shin Y-I  {\it Phys. Rev. Lett.}
{\bf 108} 035301











͑
͑

  



\end{thebibliography}
\end{document}